# A Study of Semi-Fungible Token based Wi-Fi Access Control

Litao Ye
*College of Electronics and Information Engineering*
*Shenzhen University*
Shenzhen, China
yelitao210708856@163.com

Bin Chen*
*College of Electronics and Information Engineering*
*Shenzhen University*
Shenzhen, China
bchen@szu.edu.cn
*Corresponding author

Chen Sun
*Wireless Network Research Department*
*Research and Development Center, Sony (China) Limited*
Beijing, China
chen.sun@sony.com

Shuo Wang
*Wireless Network Research Department*
*Research and Development Center, Sony (China) Limited*
Beijing, China
shuo.wang@sony.com

Peichang Zhang
*College of Electronics and Information Engineering*
*Shenzhen University*
Shenzhen, China
pzhang@szu.edu.cn

Shengli Zhang
*College of Electronics and Information Engineering*
*Shenzhen University*
Shenzhen, China
zsl@szu.edu.cn

*Abstract*—Current Wi-Fi authentication methods face issues such as insufficient security, user privacy leakage, high management costs, and difficulty in billing. To address these challenges, a Wi-Fi access control solution based on blockchain smart contracts is proposed. Firstly, semi-fungible Wi-Fi tokens (SFWTs) are designed using the ERC1155 token standard as credentials for users to access Wi-Fi. Secondly, a Wi-Fi access control system based on SFWTs is developed to securely verify and manage the access rights of Wi-Fi users. Experimental results demonstrate that SFWTs, designed based on the ERC1155 standard, along with the SFWT access right verification process, can significantly reduce Wi-Fi operating costs and authentication time, effectively meeting users' needs for safe and convenient Wi-Fi access.

## I. INTRODUCTION

With the rapid development of communication technology, Wi-Fi has become an indispensable part of modern digital life. Wi-Fi technology has evolved from the original 802.11 standard to the latest 802.11ax (Wi-Fi 6 and Wi-Fi 6E standards) [1]. Despite these advancements, Wi-Fi networks still face significant security challenges, including password cracking, man-in-the-middle attacks, and security issues in Wi-Fi access authentication and authorization [2]. Additionally, the high costs and billing difficulties associated with the centralized management and operation of public commercial Wi-Fi networks are also noteworthy [3].

The most widely used Wi-Fi security protocol is WPA2, the second-generation Wi-Fi encryption protocol released by the Wi-Fi Alliance in 2004 [4]. WPA2 is available in two versions: Home Edition and Enterprise Edition, which use preshared key (PSK) and advanced encryption standard (AES), respectively, to encrypt Wi-Fi networks. Although WPA2 provides a certain level of security for Wi-Fi networks, it also presents security risks in practical applications. For example, the key reinstallation attack (KRACK) [5], discovered by Belgian researcher Mathy Vanhoef et al. in 2017, exploits vulnerabilities in the WPA2 protocol. This attack induces users to reinstall previously used keys, enabling the attacker to crack the user's key and gain full access to the user's network. Therefore, using WPA2 in public Wi-Fi networks cannot fully guarantee user data security, quality of service (QoS), and the interests of Wi-Fi providers.

Although the Wi-Fi Alliance released WPA3 in 2018, which further improves the security and privacy protection of Wi-Fi networks [6], many user terminals and access points (APs) currently do not support the latest WPA3 protocol. Updating existing devices to support WPA3 will entail significant expenses and challenges for Wi-Fi users and providers. On the other hand, access control for APs, especially for public commercial Wi-Fi, requires batch authorization and management of a large number of users. For instance, in the traditional remote authentication dial in user service (RADIUS) authorization model [7], Wi-Fi providers usually centrally authorize and manage a large number of users simultaneously. This centralized RADIUS model is prone to single-point failures, performance bottlenecks, high management and maintenance costs, and user privacy leakage.

Blockchain is a decentralized, tamper-resistant, distributed ledger technology that enables transparent and secure transactions, along with reliable data storage. Applying blockchain technology to Wi-Fi access control may bring new possibilities for addressing network security and management overhead issues. The authors in [8] used "Colored Coins" within the Bitcoin protocol to issue Auth-Coins for Wi-Fi authentication. Nonetheless, this authentication process requires sending an authentication request to the blockchain network and waiting for confirmation. Although the security of Wi-Fi authentication is improved, the authentication speed is much slower than traditional WPA2-RADIUS. In [9], the authors used blockchain to implement anonymous access control for Wi-Fi networks. Compared to the traditional RADIUS server-based method, the proposed solution has fewer message exchanges during the authentication process and does not save sensitive user information, thereby reducing the management overhead for Wi-Fi operators. However, this authentication process also requires broadcasting the authentication transaction to the blockchain network and waiting for node verification, also resulting in a much longer authentication time compared to traditional centralized methods. In [10], authors designed a Smart Wi-Fi system that uses blockchain smart contracts and Wi-Fi billing wallets to enable Wi-Fi access through virtual currency payments and devised a Wi-Fi access fee deduction mechanism to enhance network security and user experience. Conversely, the billing mechanism of this scheme requires frequent interactions between APs and the blockchain network, increasing the operating pressure on APs and the risk of blockchain network congestion. In [11], the authors proposed N-WPA2 as a Wi-Fi authentication method using Non-Fungible Tokens (NFTs) based on ERC721 [12], effectively addressing security issues common in traditional WPA2 authentication. However, when NFTs designed based on the ERC721 standard are minted or transferred, their gas consumption increases linearly with the number of NFTs. Therefore, traditional NFTs minted based on the ERC721 standard are not suitable for Wi-Fi authorization and authentication scenarios that require large and frequent asset transfers. Additionally, the authors did not address how to implement Wi-Fi management and billing based on NFTs.

ERC1155 is a standard for creating and managing semi-fungible tokens (SFTs) [13]. Compared with single-nature tokens minted based on the ERC20 [14], ERC721, and ERC4907 [15] standards, SFTs provide a new form of digital assets that lies between fungible and non-fungible tokens. ERC1155 supports batch minting and transfer of tokens through a one-time smart contract execution. Tokens within the same set (i.e., with the same token ID) in SFTs are fungible, while tokens in different sets have unique token IDs and properties, making them non-fungible. The introduction of semi-fungible Wi-Fi tokens (SFWTs) minted based on the ERC1155 standard can provide a more personalized and flexible approach to Wi-Fi authentication and access control, enabling Wi-Fi operators to customize different types of SFWTs according to actual needs. Additionally, batch minting and transfer of SFWTs based on the ERC1155 protocol can significantly reduce the cost of executing smart contracts, thereby reducing the management complexity and cost for Wi-Fi operators while improving management efficiency. This standardized management method enables Wi-Fi operators to manage user identities and permissions more flexibly, simplify the operation process, and enhance operational efficiency and user experience.

In this work, a Wi-Fi verification process based on SFWTs is proposed to ensure the security and efficiency of Wi-Fi access. SFWTs are customized with different prices, access durations, and designated service AP information to meet the needs of Wi-Fi operators in various scenarios. The information of SFWTs is recorded on the blockchain, ensuring extremely high security. Additionally, a Wi-Fi wallet is developed, enabling users to complete verification and access Wi-Fi, while Wi-Fi operators can control and manage the Wi-Fi network.

## II. SYSTEM MODEL AND VERIFICATION PROCESS

### A. System Model

The traditional centralized WPA2-RADIUS authentication model has issues like security vulnerabilities, single-point failures, insecure user data, limited scalability, and complex management. Blockchain technology provides a decentralized alternative for verifying Wi-Fi access rights and managing operations. By using blockchain-based authentication, Wi-Fi operators can eliminate the need for centralized RADIUS servers and avoid storing sensitive user data, reducing management costs, protecting user privacy, and improving Wi-Fi network security.

A Wi-Fi access control system based on a blockchain network and SFWTs is designed. The system model is shown in Fig. 1, which mainly includes the following components:

a) Blockchain: Provides a trusted execution environment for smart contracts, securely recording, verifying, and storing user and Wi-Fi related data.

b) Wi-Fi Operators: Responsible for deploying AP devices and smart contracts, maintaining Wi-Fi wallets, uploading Wi-Fi data to the blockchain, and minting a corresponding number of SFWTs for APs based on relevant parameters.

c) User: As the end user of the Wi-Fi network, they can obtain Wi-Fi access rights by purchasing SFWTs.

d) AP: Provides wireless network connection, verifies the access rights of Wi-Fi users, and manages authorized users by querying blockchain data.

e) Minting Module: Wi-Fi operators batch mint a corresponding number of SFWTs for AP devices based on the ERC1155 protocol and Wi-Fi related parameters to achieve distributed control and billing functions of Wi-Fi networks.

f) Selling Module: Wi-Fi users can purchase SFWTs on the wallet platform using cryptocurrencies recognized by Wi-Fi operators.

g) Verification Module: Verifies whether the SFWT holder has the authority to access a specific AP and use Wi-Fi.

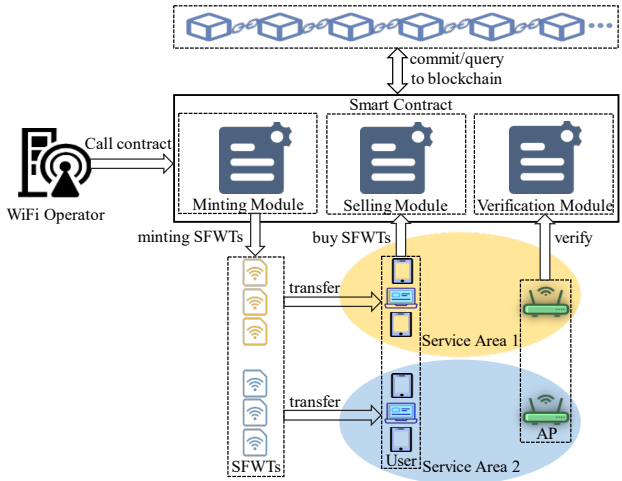

Fig. 1. System model.

## B. Verification Process

SFTs minted based on the ERC1155 standard possess characteristics of both fungible tokens (FTs) and NFTs. The ERC1155 standard supports the batch minting of tokens and assigns unique attributes to tokens with different IDs. Therefore, SFWTs are batch-minted based on the ERC1155 standard to serve as credentials for users to access Wi-Fi services. Wi-Fi operators input Wi-Fi-related parameters, such as usage time, price, available traffic, and available APs, and mint a corresponding number of SFWTs based on the AP load capacity. When users need to access Wi-Fi, they purchase the corresponding SFWTs through the Wi-Fi wallet. As shown in Fig. 2, when users obtain SFWTs and request access to Wi-Fi, they follow the Wi-Fi access verification process. This process is outlined as follows:

a) Connect to AP: The user connects to the AP they wish to access, and the user terminal displays the Wi-Fi wallet webpage. The user is currently in the *Preauthenticated* state.

b) Send ARI: The user enters their account password to log in and selects the SFWT to be verified. The user wallet sends an Authentication Request Identifier (ARI) to the AP to initiate the Wi-Fi access rights verification process.

c) Send Session ID: Upon receiving the verification identifier from the user, the AP randomly generates a 256-bit $Session_{ID}$ locally and sends it to the user.

d) User Signature: After receiving the $Session_{ID}$ from the AP, the user uses their private key ($Sk_{user}$) on the wallet to sign the $Session_{ID}$, producing the signature value $Session_{IDsig}$ = $ECDSA\_Sign(Session_{ID}, Sk_{user})$.

e) Send $Session_{IDsig}$ and SFWT ID: The user sends $Session_{IDsig}$ and the SFWT ID to the AP and waits for the verification result.

f) Verify the Signature: The AP uses the $Session_{ID}$ and $Session_{IDsig}$ to recover the signer address ($Addr_{user}$) based on the ECDSA algorithm: $Addr_{user} = ECDSA\_Recover(Session_{IDsig}, Session_{ID})$.

g) Access Authority Verification: The AP uses $Addr_{user}$, SFWT ID, and the amount of data ($usedData_{user}$) to query the blockchain network to verify the user's access authority. Upon successful verification, the AP adds or updates $Addr_{user}$ and the remaining access time to the local authorized user list.

h) Return Verification Results: The AP returns the verification results and the remaining Wi-Fi service time to the user's Wi-Fi wallet. Upon successful verification, the user is in the *Authenticated* state and can use the Wi-Fi service provided by the AP.

i) Disconnect: The AP periodically checks the local authorized user list and re-verifies users with expired services to ensure they no longer have access rights. It removes the MAC address of users whose verification has expired, thereby terminating their Wi-Fi service.

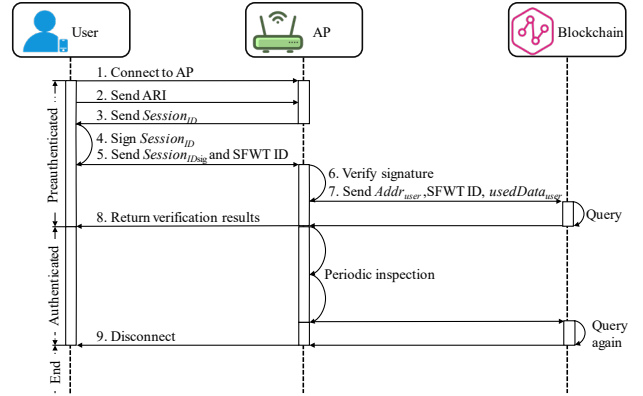

Fig. 2. Verification process.

## III. SFWT BASED ON ERC1155

### A. Minting of SFWTs

The ERC1155 standard provides the functionality for batch minting and transferring tokens. With a single interaction with the blockchain, both batch minting and transfer of tokens can be accomplished, significantly reducing gas costs associated with minting and transferring SFWTs. For Wi-Fi providers, this translates to substantial savings in operating costs, allowing them to offer better QoS and improve system efficiency. Consequently, SFWTs are minted using the ERC1155 standard as per the system model outlined in section 2.1. Smart contract deployment and execution are carried out on the Ethereum network.

Algorithm 1 presents the pseudocode for the minting contract. In this contract, the Wi-Fi provider generates SFWTs with identical IDs by inputting parameters such as the Wi-Fi provider address, AP ID, price, usable time, usable data volume, and minting quantity.

**Algorithm 1** Smart Contract (mint SFWTs)
---
**Require:** *ERC-1155 of openzeppelin*
**Input:** owner, tokenId, APId, price, duration, dataCap, quantity
**Output:** boolean
1: **if** (caller = WiFi operator or administrator) **then**
2:   owner ← WiFi operator address
3:   tokenId ← SFWT ID
4:   _mint(owner, tokenId, quantity) //ERC-1155 internal function
5:   SFWTMetadata[tokenId].APId ← APId
6:   SFWTMetadata[tokenId].price ← price
7:   SFWTMetadata[tokenId].duration ← duration
8:   SFWTMetadata[tokenId].dataCap ← dataCap
9:   **return** true
10: **else return** false

Fig. 3 illustrates an example result of a minting operation, where the SFWT ID is 1, the minting quantity is 10, and the token owner address is "0xa8126934 003110d5b7eC9a275e2 7B6d2fFA28529."

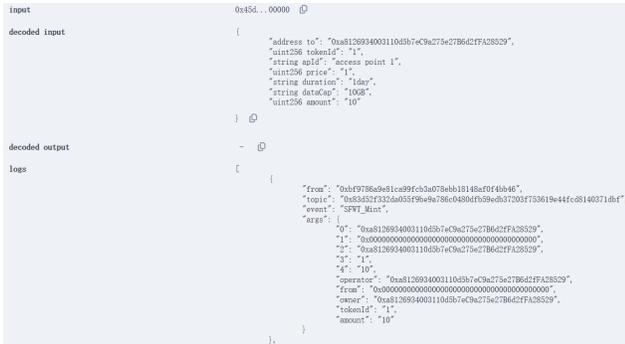

Fig. 3. SFWTs batch minting results.

### B. Purchase and Verification of SFWT

After the user pays a certain amount of ETH (Ethereum network native currency) to the Wi-Fi operator to purchase SFWTs, they attempt to connect to a specific AP deployed by the Wi-Fi operator. When the user attempts to access, the Wi-Fi wallet webpage will automatically pop up on the user's terminal. The user needs to enter the corresponding account password on this page to access their personal account. Subsequently, the user uses the SFWT that meets the requirements to complete the Wi-Fi access permission authentication on the blockchain network.

Algorithm 2 shows the pseudocode for the purchase and verification of SFWT.

**Algorithm 2** Smart Contract (buy and verify SFWT)
---
**Require:** *ERC-1155 of openzeppelin*
**Part 1: buy SFWT**
**Input:** tokenId, quantity, sum
**Output:** boolean
1: **if**(sum = SFWTMetadata[tokenId].price * quantity)
2:   payable(owner).transfer(sum)
3:   _safeBatchTransferFrom(owner, caller, tokenId, quantity) //ERC-1155 internal function
4: **if**(currentTime >= WiFiInfo[tokenId].expirationTime[caller]) **then**
5:   expirationTime ← currentTimestamp + SFWTMetadata[tokenId].duration * quantity
6:   WiFiInfo[tokenId].expirations[caller] ← expirationTime
7: **else if**(currentTime < WiFiInfo[tokenId].expirationTime[caller]) **then**
8:   WiFiInfo[tokenId].expirations[caller] += SFWTMetadata[tokenId].duration * quantity
9: **return** true
10: **else return** false
**Part 2: verify SFWT**
**Input:** tokenId, currentAPId
**Output:** boolean
1. **if**(balanceOf(caller, tokenId) > 0) **then**
2.   **if**(currentAPId = SFWTMetadata[tokenId].APId) **then**
3.     **if**(currentTime < WiFiInfo[tokenId].expirationTime[caller]) **then**
4.       **if**(amountOfDataUsed < SFWTMetadata[tokenId].dataCap * balanceOf(caller, tokenId))
5.         **then return** true
6. **else return** false

## IV. EXPERIMENTAL RESULTS AND ANALYSIS

### A. Experimental Process

To improve the Wi-Fi user experience and simplify the management process, a Wi-Fi wallet was developed based on Vue.js and Web3.js libraries. The wallet allows users to manage assets and authenticate identities on a web page. Fig. 4 shows the Wi-Fi wallet client details, which lists all SFWTs owned by the user and their related information, as well as other functions.

Given the Raspberry Pi's user-friendly development features and high scalability, it was utilized as an AP in this experiment. Its Wi-Fi-related parameters were configured, the WPA2 verification function was deactivated, and Nodogsplash [16], which provides a mandatory portal function, was employed as the mandatory verification tool for Wi-Fi access. As shown in Fig. 4, among the SFWTs, the token with ID 1 cannot be used for Wi-Fi access because it has expired. The token with ID 2 is a valid token and is used to connect to the AP with ID AP1.

When the user selects the SFWT with ID 2 as the credential to access this AP, the user, the AP, and the blockchain network execute the user's access authority verification according to the verification interaction process designed in section 2.2. Upon successful verification, the AP returns the verification result and displays the remaining connection time, remaining available data, and other related information on the Wi-Fi wallet at the user's equipment. At this point, the user can use the network services provided by the AP.

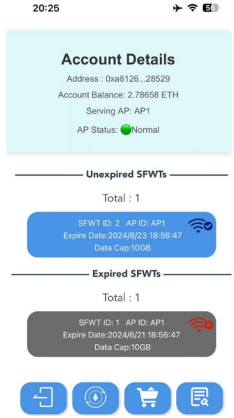

Fig. 4. Wi-Fi wallet client details.

*B. Performance Analysis*

The ERC721 and ERC1155 standards were used to operate (mint/transfer) Wi-Fi tokens in the same Ethereum network environment, ensuring consistent token parameters. Fig. 5 compares the gas consumption of the two standards when operating different numbers of Wi-Fi tokens per transaction. Gas is the fee required to execute transactions or smart contracts in the Ethereum network, reflecting the resource consumption and efficiency of the operation. The experiment was conducted 100 times, and the results indicated that the gas consumption of the ERC721 and ERC1155 standards was essentially the same when operating a single Wi-Fi token. However, when minting and transferring a large number of Wi-Fi tokens, the gas consumption of the ERC721 standard increased linearly, whereas the gas consumption of the ERC1155 standard was unaffected by the number of tokens. Therefore, this scheme exhibits low cost and low complexity when minting and transferring SFWTs, making it highly suitable for bulk authorization and management of Wi-Fi tokens.

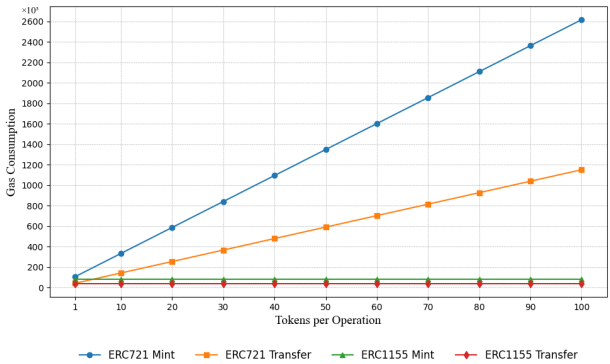

Fig. 5. Gas consumption for various Wi-Fi token operations.

In addition, the verification time of different Wi-Fi access authentication schemes was compared and analyzed. Existing Wi-Fi authentication schemes were classified into three categories: the traditional WPA2 scheme, the block broadcast verification transaction scheme (see references [8][9][10]), and the N-WPA2 authentication scheme in reference [11]. A verification time comparison experiment was conducted on the scheme proposed in this paper and the existing schemes in the same environment.

The experiment was performed on the same AP and user terminal, with the block generation time interval set to 10 seconds in the Ethereum network environment. For the aforementioned verification methods, 100 experiments were conducted for each method, excluding the time for the user to enter the password. Fig. 6 illustrates the comparison results of Wi-Fi authentication time across different schemes.

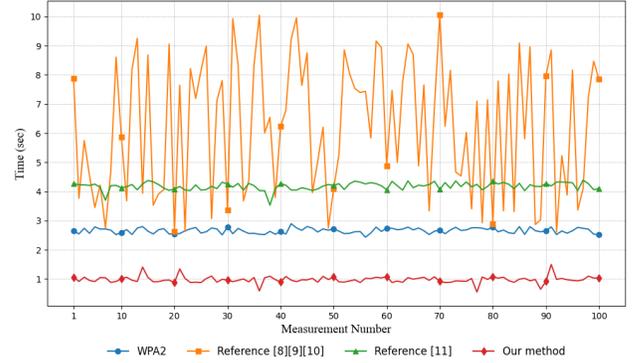

Fig. 6. Comparison of Wi-Fi authentication time for different solutions.

In the comparison, the block broadcast verification transaction scheme exhibited the largest verification time fluctuation due to the need to wait for the transaction to be packaged and broadcasted by the block, significantly affected by block generation speed and network congestion. The N-WPA2 scheme requires NFT information query verification before WPA2 verification. Although the information query time on the blockchain is generally stable, the scheme's need to be compatible with existing APs necessitates performing the AP built-in WPA2 authentication process after the NFT query verification, resulting in longer verification time than direct WPA2 verification.

The proposed verification scheme utilizes Nodogsplash as the verification portal and can operate on OpenWrt and other Linux-based systems. Given that most router systems are customized based on these systems, the scheme is compatible with most existing APs and only requires corresponding software upgrades for the AP. The proposed scheme does not necessitate the WPA2 verification process, but only requires querying and verifying SFWT on the blockchain. This query request is minimally affected by blockchain network congestion and does not require executing transactions and waiting for them to be packaged and confirmed by blocks. Consequently, the scheme significantly reduces verification time while providing stability comparable to WPA2.

V. CONCLUSION

A Wi-Fi access authentication scheme based on blockchain smart contract technology is proposed to address the challenges of security issues and high management costs inherent in traditional Wi-Fi access authentication models. The ERC1155 standard is used to mint SFWTs as credentials for users to access Wi-Fi. Users can purchase SFWTs through Wi-Fi wallets, complete Wi-Fi access authentication, and use Wi-Fi networks.

The minting, trading, and verification processes of SFWTs are all implemented through blockchain smart contracts. Experimental results demonstrate that the proposed scheme exhibits strong scalability, low cost, and short verification time.

ACKNOWLEDGMENT

Guangdong Province Graduate Education Innovation Program Project[2024JGXM_163], Shenzhen University Graduate Education Reform Research Project [SZUGS2023JG02], Shenzhen University Teaching Reform Research Project [JG2023097], Foundation of Shenzhen [20220809155455002, 20220810142731001], National Natural Science Foundation of China under grant 62171291, Shenzhen Key Re-search Project[JSGG20220831095603007, JCYJ20220818100810023, JCYJ20220818101609021]. Shenzhen University High-Level University Construction Phase III -Human and Social Sciences Team Project for Enhancing Youth Innovation[24QNCG06].